\documentclass[a4paper,11pt]{article}
\usepackage{jcappub} 
\usepackage{lineno}
\usepackage{graphicx}
\usepackage{amsmath}
\usepackage[nopatch]{microtype}
\usepackage{booktabs}
\usepackage{hyperref}
\usepackage{url}
\usepackage{graphics}
\usepackage{xurl}
\usepackage{balance}
\usepackage{soul}
\usepackage[T1]{fontenc}
\usepackage{xspace}
\usepackage{tikz}
\usepackage{amsmath,amssymb}
\usepackage{bm}
\usepackage{cancel}
\usepackage[normalem]{ulem}
\newcommand{\hcon}{\ensuremath{\mathrm{km~s^{-1}~Mpc^{-1}}}\xspace}

\usepackage{natbib}
\usepackage{booktabs}

\title{\boldmath Dark siren cross-correlations and the sensitivity of $H_0$ to methodological choices}

\author[a,c]{Madeline L. Cross-Parkin,}
\author[a,c]{Cullan Howlett,}
\author[b,c]{Leonardo Giani,}
\author[b,c]{Chris Blake,}
\author[a,c]{Tamara M. Davis,}

\affiliation[a]{School of Mathematics and Physics, University of Queensland, Brisbane, Australia}
\affiliation[b]{Swinburne University of Technology, Hawthorn VIC 3122, Australia}
\affiliation[c]{OzGrav: The ARC Centre of Excellence for Gravitational Wave Discovery}

\emailAdd{m.crossparkin@uq.edu.au}
\emailAdd{c.howlett@uq.edu.au}
\emailAdd{lgiani@swin.edu.au}
\emailAdd{cblake@swin.edu.au}
\emailAdd{tamarad@physics.uq.edu.au}

\abstract{Gravitational wave sources act as absolute distance indicators, making them powerful probes of the present-day expansion rate of the Universe, $H_0$. The cross-correlation method combines gravitational wave events with galaxy catalogues to constrain cosmological parameters through their shared large-scale structure. In this work, we investigate how key methodological choices---including covariance treatment, bias parametrisation for galaxies and gravitational wave events, and distance and redshift binning width---affect the inferred value of $H_0$. We also study catalogue incompleteness, showing that selection effects can be incorporated directly into the theoretical prediction, without the need to model the missing population explicitly, a key advantage over the standard galaxy catalogue approach. Our results indicate that, with appropriate modelling choices and a sufficiently large sample of precise gravitational wave events, the systematic biases considered here can be effectively mitigated, highlighting the potential of the cross-correlation method for future dark siren precision cosmology.}

\begin{document}
\maketitle

\flushbottom

\section{Introduction}

Modern cosmology is increasingly driven by detailed measurements of the Universe’s large-scale structure, which provide a powerful means of probing cosmic evolution and testing the standard $\Lambda$CDM model. Persistent tensions in parameters inferred within this concordance framework---most notably in measurements of the Hubble constant, $H_0$ \citep{Verde_2019}---motivate the quest for innovative and independent avenues to measure cosmological parameters. One such avenue, made possible by the first gravitational wave detections by the LIGO-Virgo-KAGRA (LVK) Collaboration \citep{2015CQGra..32g4001L,2015CQGra..32b4001A, 2019}, is the field of gravitational wave cosmology. Gravitational waves act as absolute distance indicators, making them powerful cosmological probes known as standard sirens \citep{Schutz}.

Several techniques have been developed to use gravitational waves to measure $H_0$, using either \textit{bright sirens} or \textit{dark sirens}. In the former case, the identification of an electromagnetic counterpart to the gravitational wave event enables a direct measurement of the source redshift, allowing for an inference of $H_0$ through the luminosity distance--redshift relation \citep{2017Natur.551...85A}. In the latter, the redshift is inferred statistically by associating the gravitational wave localisation area with galaxies in existing catalogues. These redshifts are then combined with the inferred luminosity distance to construct a Hubble diagram and constrain $H_0$ \citep{Chen:2017rfc, Finke:2021aom, 2023AJ....166...22G, Gray:2019ksv, Holz_2005, LIGOScientific:2018gmd, LIGOScientific:2021aug, PhysRevD.86.043011, Soares_Santos_2019, theligoscientificcollaboration2025gwtc40constraintscosmicexpansion, mcmahon2026measurementhubbleconstantusing,alfradique2026improvedconstrainthubbleconstant,theligoscientificcollaboration2026gwtc50constraintscosmicexpansion}. Another complementary approach, known as the spectral siren method \citep{farah2024needknowastrophysicsfreegravitationalwave}, exploits our understanding of the black hole mass spectrum to effectively “calibrate” gravitational wave events, allowing their redshifts to be inferred from the observed chirp mass, $\mathcal{M}_z=\mathcal{M}(1+z)$. This is often used in conjunction with dark siren methods, where it helps break the mass--redshift degeneracy and thereby improves constraints on $H_0$.

While both bright and dark siren methodologies have improved considerably, there remains significant uncertainty arising particularly from incomplete galaxy catalogues and assumptions underlying the modelling of compact object population properties. While many of these systematic effects can be accounted for---and substantial progress has been made in the literature to do so (see \citet{Gray_2022, Borghi_2026})---there remains a valuable opportunity to explore complementary dark siren methodologies.

In this work, we explore an alternative dark siren approach to cosmological inference based on the clustering of gravitational wave sources. As tracers of the underlying dark matter distribution, gravitational wave events are spatially correlated with other large-scale structure probes, particularly galaxies. These correlations can be used to extract cosmological information, as well as to infer properties such as the redshift distribution of the sources. This has been demonstrated in applications to photometric samples in gravitational lensing surveys \citep{2008ApJ...684...88N}. \citet{Oguri_2016} first proposed that the clustering of gravitational wave sources in luminosity distance space, together with the clustering of galaxies in redshift space, can be used to constrain the redshift--distance relation. While most current literature uses the 2D angular power spectrum to probe this clustering, \citet{Ghosh_2025} used the 3D cross-correlation function. This approach has the advantage of exploiting both radial and angular clustering, thereby retaining the full three-dimensional modes rather than projecting the signal into two dimensions. However, a full 3D cross-correlation analysis requires careful modelling of redshift-space distortions and becomes increasingly difficult to model in the presence of large observational uncertainties. In contrast, angular localisation uncertainty can be incorporated naturally through a window function $W_\ell$, which suppresses power on small angular scales (high $\ell$). This formulation allows detector effects to be modelled in a straightforward and computationally efficient manner.

The angular cross-correlation method has been used to provide forecasts on cosmological parameters with both future galaxy surveys and next-generation detectors \citep{Pan:2025iya,Ferri_2025, pedrotti2025cosmologyangularcrosscorrelationgravitationalwave, Calore_2020}. This technique has been recently applied to real gravitational wave data in \citet{Mukherjee_2024} and \citet{dematos2025measurementhubbleconstantgravitational}, with very promising results. In addition, \citet{sala2025inferringcosmologicalparametersgalaxy} forecasts how future detectors and surveys will constrain cosmological parameters using the cross-correlation method. However, as is well known in large-scale structure cosmology, such innovative approaches inevitably contend with a complex landscape of observational and systematic biases, alongside numerous methodological choices.

The aim of this paper is to examine the methodological choices inherent to the cross-correlation method and their impact on the inferred value of $H_0$. The structure of the paper is as follows. In Section~\ref{sec:theory}, we briefly review the theory of angular cross-correlations and introduce the methodology. In Section~\ref{sec:results}, we present the main results of this work, examining the performance of the cross-correlation method under a range of methodological choices and systematic effects, and quantify their impact on the inferred value of $H_0$. Specifically, Section~\ref{sec:sijvscl} examines the impact of compressing the cross-power spectra into a single representative statistic, and evaluates the extent to which this compression preserves the cosmological information relevant for constraining $H_0$. Section~\ref{subsec:cov} compares analytic covariance estimates with those derived from a suite of log-normal mocks, highlighting the implications for parameter inference. Section~\ref{sec:systematics} investigates the impact of key systematic uncertainties on the cross-correlation method, including Malmquist and Eddington biases, the effects of incorrect bias parametrisation, and biases arising from catalogue incompleteness. Finally, Section~\ref{subsec:binning} explores how the choice of tomographic binning in redshift and luminosity distance influences the uncertainty on $H_0$, and how appropriate binning strategies can help mitigate the impact of large observational uncertainties. In Section~\ref{sec:discussion}, we summarise our findings and discuss prospects for future work.

\section{Formalism and methodology}
\label{sec:theory}
Both galaxies and gravitational wave events trace the underlying matter distribution, and thus cluster within the same large-scale dark matter inhomogeneities. The overlap in their clustering can be quantified by their angular cross-correlation, which measures the excess probability of finding a galaxy and a gravitational wave event at a given angular separation.

While gravitational wave signals directly encode the luminosity distance to their sources, galaxy redshifts can only be converted into distances by assuming a cosmological model. The cross-correlation method exploits this by identifying the cosmological parameters that maximise the angular cross-correlation between gravitational wave events and galaxies once their redshifts are mapped to distances.

\subsection{Angular cross-correlation}
\label{sec:angularcrosscorrelation}

From the observed angular distribution of tracer
$X$, we can define the angular density contrast field
\begin{equation}
    \delta^{X}(\theta,\phi) \equiv \frac{n_X(\theta,\phi)}{\bar n_X} - 1,
\label{eq:delta_map}
\end{equation}
where $\bar n_X$ is the mean angular number density of tracer $X$. We can then expand $\delta^X$ in a basis of spherical harmonics
\begin{equation}
    \delta^{X}(\hat{\mathbf{n}})=\sum_{\ell m} a_{\ell m}^X\,Y_{\ell m}(\hat{\mathbf{n}}),
\end{equation}
where the coefficients $a_{\ell m}^X$ encode the angular clustering information of tracer $X$.

Introducing a second tracer $Y$, we can similarly define $\delta^Y$ and its harmonic coefficients $\alpha^Y_{\ell m}$. Under the assumption of statistical isotropy, the joint two-point statistics of the two tracers define the angular cross-power spectrum,
\begin{equation}
    \left\langle a_{\ell m}^{X}\;,\; a_{\ell' m'}^{Y^*} \right\rangle=        \delta_{\ell\ell'}\,\delta_{mm'}\,C_\ell^{XY}.
\label{eq:cell_def}
\end{equation}

Working in Fourier space and projecting the three-dimensional fluctuations onto the sphere, the angular cross-power spectrum between two tracers, $X$ and $Y$, can be written following e.g. \citet{Giani_2023} and \citet{Sch_neberg_2018} as

\begin{equation}
    C_\ell^{XY}=\frac{2}{\pi}\int_0^\infty dk\,k^2 P(k)\,\Delta^{X}_{\ell}(k)\,\Delta_{\ell}^{Y}(k)\;,
\label{eq:cell_kernel}
\end{equation}
where we have defined the projection kernel for tracer $X$ as
\begin{equation}
    \Delta^X_\ell(k)=\int dz\,W_X(z)\,b_X(z)\,D(z)\,j_\ell\!\left(k\,\chi(z)\right)\;,
\label{eq:kernel_def}
\end{equation}
with an analogous expression holding for tracer $Y$. Here, $D(z)$ is the linear growth factor, $j_\ell$ are spherical Bessel functions, $b_X(z)$ is the bias of tracer $X$, and $W_X(z)$ denotes the window function describing the radial selection of the tracer. For a full-sky homogeneous distribution, this window function is given by the normalised redshift distributions,
\begin{equation}
    W_X(z) = \frac{\bar{n}_X(z) \chi^2(z)}{H(z) N}\;, \qquad N=\int dz \frac{\bar{n}_X(z) \chi^2(z)}{H(z)}\;,
\end{equation}
where $\bar{n}$ is the comoving number density of the tracer $X$ and $N$ ensures proper normalisation within each redshift bin. Additionally, $\chi(z)$ and $H(z)$ are the comoving distance and Hubble parameter, respectively.

To reduce noise and improve numerical stability, it is convenient to work with binned angular power spectra, defined as
\begin{equation}
    {C}_{l_b}^{XY}=\sum_{\ell \in l_b} W_{\ell}\, C_\ell^{XY},
\label{eq:binned_pseudo_cl}
\end{equation}
where \(W_{l_b}\) denotes the binning operator. 

In practice, incomplete sky coverage and selection effects lead to a masked overdensity field $\tilde{\delta}^X$, from which a pseudo-angular power spectrum $(\tilde C_\ell^{XY})$ is measured. The expectation value of the pseudo-power spectrum is related to the underlying true angular power spectrum through mode coupling induced by the survey mask
\begin{equation}
\left\langle \tilde C_\ell^{XY} \right\rangle
=
\sum_{\ell'}
M_{\ell\ell'}^{XY}\, C_{\ell'}^{XY},
\label{eq:master}
\end{equation}
where \(M_{\ell\ell'}^{XY}\) is the mode-coupling matrix determined by the geometry of the mask.

In this work, we use binned angular power spectra and perform mask deconvolution using the \texttt{NaMaster} package\footnote{\href{https://namaster.readthedocs.io/en/latest/}{https://namaster.readthedocs.io/en/latest/}} \citep{Alonso_2019}, which implements the MASTER formalism to provide unbiased estimates of the underlying angular power spectrum.

\subsection{Tomographic angular cross-correlation of gravitational waves and galaxy catalogues}
\label{sec:tomographicangular}

The cross-correlation method is typically implemented within a tomographic framework, as the angular power spectrum is inherently a two-dimensional statistic. Tomographic binning in redshift and luminosity distance is therefore required to retain sensitivity to the radial distribution of gravitational wave sources and to extract cosmological information along the line of sight. 

We consider gravitational wave events with observed luminosity distances in the bin \(d_{L,j} < \hat{d}_L < d_{L,j+1}\), and galaxies with observed redshifts in the bin \(z_i < \hat{z} < z_{i+1}\). Throughout this work, observed quantities are denoted with hats \((\hat{\ })\) to indicate that they are subject to measurement uncertainty. The angular cross-power spectrum for a given \((i,j)\) bin pair can then be written as
\begin{equation}
\label{eq:Clij}
C_{\ell}^{\text{gal}_{i}\times\text{GW}_{j}} = \frac{2}{\pi}\int_{0}^{\infty}dk\,k^{2} P(k) \Delta_{\ell}^{\text{gal}_i}(k)\Delta_{\ell}^{\text{GW}_j}(k)
\end{equation}
where $\Delta_{\ell}^{\text{gal},i}$ and $\Delta_{\ell}^{\text{GW}, j}$ are the projection kernels for galaxies and gravitational wave events in bins $i$ and $j$, respectively. These kernels encode the effects of selection, binning, and observational uncertainties. We follow a formulation similar to \citet{sala2025inferringcosmologicalparametersgalaxy} and \citet{pedrotti2025cosmologyangularcrosscorrelationgravitationalwave}, writing the kernels as

\begin{align}
\label{eq:kernel1}
    \Delta_{\ell}^{\text{gal}_i} &= \frac{1}{N^{\text{gal}_i}}\int_{0}^{\infty}dz\,\frac{\chi^{2}(z)}{H(z)}P(\hat{z} \in \mathrm{bin\,}i|z, \mathrm{selection})b_{\text{gal}}(z)D(z)j_{\ell}(k\chi(z)), \\
    \label{eq:kernel2}
    \Delta_{\ell}^{\text{GW}_j} &= \frac{1}{N^{\text{GW}_j}}\int_{0}^{\infty}dz\,\frac{\chi^{2}(z)}{H(z)}P(\hat{d_L} \in \mathrm{bin\,}j|z, \mathrm{detection})b_{\text{GW}}(z)D(z)j_{\ell}(k\chi(z)).
\end{align}
Here, $N^{\mathrm{gal}_i}$ and $N^{\mathrm{GW}_j}$ are normalisation factors for each bin, $b_{\mathrm{gal}}(z)$ and $b_{\mathrm{GW}}(z)$ denote the galaxy and gravitational wave biases.

The remaining non-trivial components in Equations~\ref{eq:kernel1} and \ref{eq:kernel2} encode the probabilities of observing a redshift $\hat{z}$ or distance $\hat{d}_L$ to be in a bin $i$ or bin $j$ given the true unknown redshift of the event, our measurement uncertainty and any selection or detection functions. For gravitational waves, in this work we assume a Gaussian likelihood for \(\hat{d}_L\) about the true luminosity distance \(d_L(z)\), with uncertainty \(\sigma_{d_L}\), and impose an upper limit $d_L^{\rm thr}$. The corresponding probability is then
\begin{align}
P(\hat{d}_L \in \mathrm{bin}\ j \mid z, \mathrm{detection})
&=
\int_{d_{L,j}}^{d_{L,j+1}}
\mathcal{N}\!\left(\hat{d}_L \mid d_L(z), \sigma_{d_L}\right)
P(\hat{d}_L \mid d_L^{\rm thr})\,{\rm d}\hat{d}_L \nonumber \\
&=
\int_{d_{L,j}}^{\min(d_L^{\rm thr},\,d_{L,j+1})}
\mathcal{N}\!\left(\hat{d}_L \mid d_L(z), \sigma_{d_L}\right)
\,{\rm d}\hat{d}_L \nonumber\\
&=
\Phi\!\left(
\frac{\min(d_L^{\rm thr},\,d_{L,j+1}) - d_L(z)}
{\sigma_{d_L}}
\right)
-
\Phi\!\left(
\frac{d_{L,j} - d_L(z)}
{\sigma_{d_L}}
\right),
\label{eq:probdl}
\end{align}
where $\Phi$ denotes the cumulative distribution function of the Gaussian distribution. 

Similarly, we account for redshift uncertainties by incorporating them into the radial selection through the probability $P(\hat{z} \in \mathrm{bin}\, i \mid z)$. This quantity represents the probability that a galaxy with true redshift $z$ is assigned an observed redshift $\hat{z}$ within the bin $z_i < \hat{z} < z_{i+1}$. Assuming a Gaussian redshift likelihood with uncertainty $\sigma_z$, and a selection function that depends only on the true redshift, this probability can be written in separable form. This approximation is appropriate for simple selections, such as a magnitude limit modelled through the luminosity function and true redshift. In this case,
\begin{equation}
\begin{split}
    P(\hat{z} \in \mathrm{bin}\ i \mid z, \mathrm{selection})
    &= P(\hat{z} \in \mathrm{bin}\ i \mid z)\,P_{\mathrm{selection}}(z) \\
    &= \left[
    \Phi\left(\frac{z_{i+1}-z}{\sigma_z}\right)
    - \Phi\left(\frac{z_i-z}{\sigma_z}\right)
    \right] P_{\mathrm{selection}}(z)\;.
\end{split}
\label{eq:probz}
\end{equation}
This separability may not hold for more complex observational effects, such as spectroscopic redshift success, which can depend on additional galaxy properties and observing conditions. In the limiting case of a fully complete catalogue, where no selection effects are present, $P_{\mathrm{selection}}(z) = 1$.

Combined with the volumetric factor $\chi^{2}(z)/H(z)$ and the normalisation terms $N^{\text{GW}_j}$ and $N^{\text{gal}_i}$, these selection probabilities can be absorbed into normalised redshift distributions for each tracer. We therefore define the effective redshift distributions $dn^{\mathrm{gal}_i}/dz$ and $dn^{\mathrm{GW}_j}/dz$, which describe the number of galaxies and gravitational wave events per unit redshift in bins $i$ and $j$, respectively. In terms of these quantities, the kernels take the simplified form

\begin{align}
    \label{eq:kernel1simp}
    \Delta_{\ell}^{\text{gal}_i} &= \int_{0}^{\infty}dz\,\frac{dn^{\text{gal}_i}}{dz}b_{\text{gal}}(z)j_{\ell}(k\chi(z)),\\
    \label{eq:kernel2simp}
    \Delta_{\ell}^{\text{GW}_j} &= \int_{0}^{\infty}dz\,\frac{dn^{\text{GW}_j}}{dz}b_{\text{GW}}(z)j_{\ell}(k\chi(z)).
\end{align}

\begin{figure}
    \centering
    \includegraphics[width=0.85\linewidth]{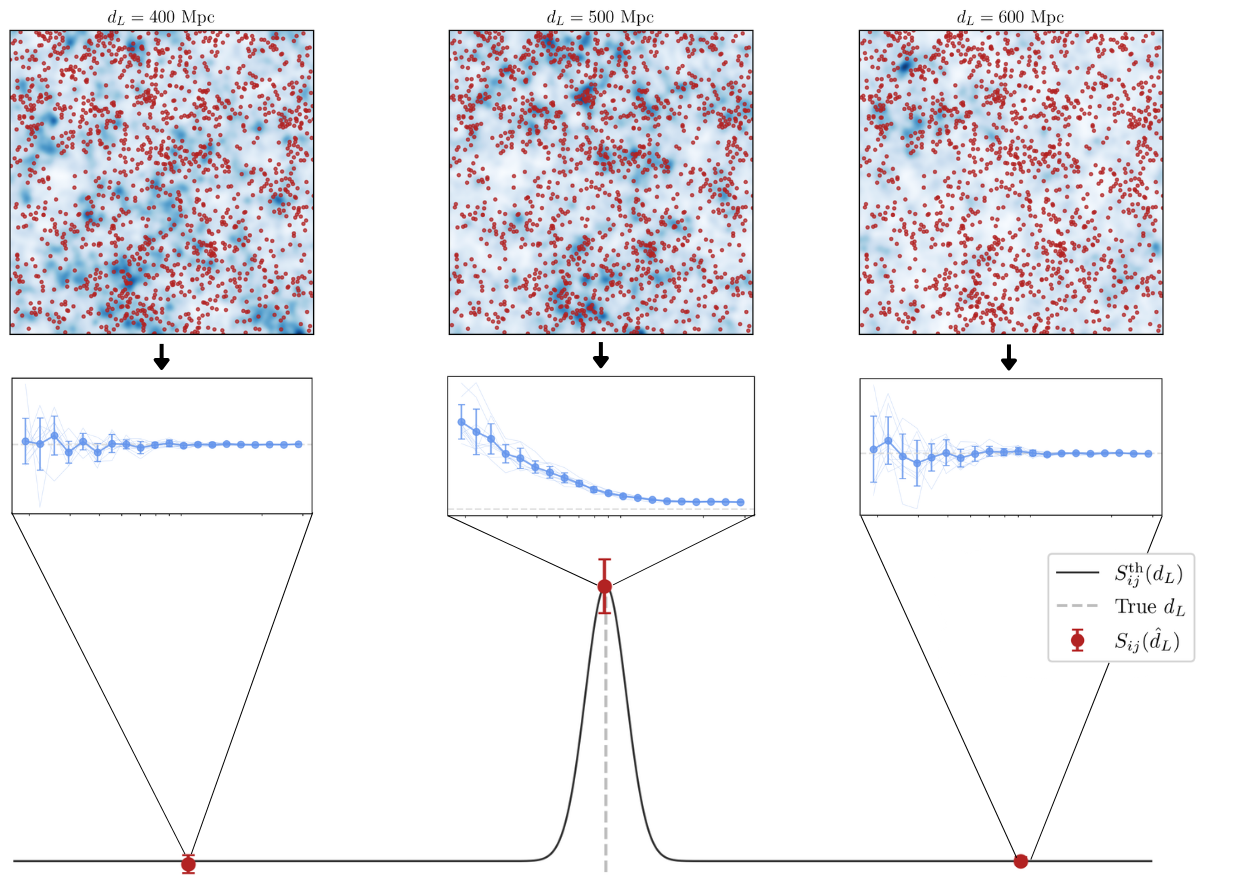}
    \caption{Schematic illustration of the cross-correlation method. The upper panels show the projected galaxy distribution (red points) and gravitational wave probability field (blue shading) in three luminosity distance slices. For each slice, the angular cross-correlation between the two fields is measured as a function of angular scale, shown by the blue curves and points in the inset panels. The black curve in the lower panel shows the theoretical cross-correlation signal, $S_{ij}^{\rm th}(d_L)$, as a function of luminosity distance, while the red points indicate the measured statistic, $S_{ij}(\hat{d}_L)$, obtained by taking a weighted sum of the angular cross-correlation over $\ell$. The signal peaks when the luminosity distance slice gives the strongest overlap between the gravitational wave and galaxy distributions, allowing the true distance--redshift relation to be inferred. Adapted from \citet{Bera_2020}, with the separation set to $r=0$ in the configuration-space cross-correlation function.}
    \label{fig:clinset}
\end{figure}

The strength of the angular cross-correlation is governed by the overlap between the gravitational wave and galaxy tracer populations. We therefore characterise the correlation amplitude using a single summary statistic derived from the angular power spectrum. Specifically, we define the cross-correlation compression
\begin{equation}
\label{eq:sij}
    S_{ij} \equiv \sum_{\ell}^{\ell_{\mathrm{max}}} \frac{2\ell+1}{4\pi}C_{\ell}^{\mathrm{gal}_i\times\mathrm{GW}_j},
\end{equation}
which provides a convenient, scalar measure of the total cross-correlation strength between gravitational wave events in luminosity distance bin $j$ and galaxies in redshift bin $i$. This $S_{ij}$ parameter is illustrated in Figure~\ref{fig:clinset}. A detailed discussion of the advantages and limitations of this compressed statistic, relative to the full power spectra, is presented in Section~\ref{sec:sijvscl}, while the physical interpretation of this compressed statistic is explored in Appendix~\ref{app:theorysij}.

\subsection{Parameter inference and pipeline}
\label{sec:pipeline}
We measure $S_{ij}$ in tomographic galaxy redshift bins, evaluating the cross-correlation strength in each bin as a function of gravitational wave luminosity distance across a set of distance bins. Collecting all $(i,j)$ measurements into a single data vector $\mathbf{S}$, the likelihood can be written as

\begin{align}
\label{eq:likelihood}
\ln \mathcal{L}(\bm{\theta})
=
-\frac{1}{2}
\left(\mathbf{S}^{\rm data}-\mathbf{S}^{\rm th}(\bm{\theta})\right)^{\rm T}
\mathbf{\Sigma}^{-1}
\left(\mathbf{S}^{\rm data}-\mathbf{S}^{\rm th}(\bm{\theta})\right),
\end{align}
where bold symbols denote vectors constructed from all $(i,j)$ bins. The covariance matrix $\mathbf{\Sigma}$ can be defined both analytically or using mocks. We discuss and compare these two prescriptions in Section~\ref{subsec:cov}. 
The parameter vector is given by $\bm{\theta} = \{H_0, \mathbf{a}^{\rm gal}, \mathbf{b}^{\rm GW}\}$, and the theoretical prediction is given by
\begin{align}
    \mathbf{S}^{\rm th}(\bm{\theta}) =A_{ij}(z)S_{ij}^{\rm th}(z_i,d_{L,j}\mid H_0, b(z)=1).
\end{align}
Here, $A_{ij}(z)$ encodes the effective bias of the galaxy and gravitational wave tracers and is defined as a kernel-weighted average of the bias product
\begin{equation}
\label{eq:Aij}
A_{ij}(z)
=
\frac{
\displaystyle
\int dz \;
K_{ij}(z)\,
b_{\rm GW}(z)\,
b_{\rm g}(z)
}{
\displaystyle
\int dz \;
K_{ij}(z)
},
\end{equation}
\begin{equation}
    K_{ij}(z)=\frac{dn^{\text{GW}_i}}{dz}\frac{dn^{\text{gal}_j}}{dz}\;,
\end{equation}
We stress that $A_{ij}(z)$ is not, in general, equal to the product of the biases evaluated at a single redshift. Rather, it reduces to it only in the limit where the redshift bin is sufficiently narrow or where the kernel $K_{ij}(z)$ is sharply peaked such that the gravitational wave and galaxy windows maximally overlap at a single effective redshift.

Finally, $S_{ij}^{\rm th}$ denotes the baseline theoretical prediction, computed using \texttt{CAMB} \citep{2011ascl.soft02026L} under the assumption of unit bias. Some studies (for example \citet{Bera_2020}) fit a Gaussian profile to the binned cross-correlation function in real space and use the peak location to infer $H_0$. In the course of our investigations, we found that this approach implicitly absorbs galaxy/gravitational wave bias effects into the fitted profile, rather than explicitly modelling their redshift dependence. In doing so, it compresses the information contained in this statistic into a single parameter---the mean of the Gaussian---thereby discarding additional information encoded in higher-order moments of the signal. The likelihood for $S_{ij}$ built in Equation~\ref{eq:likelihood} is designed to capture this additional information.  

\subsection{Data and simulations}
We generate a suite of 600 log-normal mock catalogues based on a flat $\Lambda$CDM cosmology with parameters $H_0=70$ \hcon, $\Omega_m=0.25$, $\Omega_b=0.044$, and $\Omega_\Lambda=0.75$, spanning the redshift range $0 \leq z \leq 0.5$. These mocks are constructed by first generating a Gaussian random field and then applying a log-normal transformation \citep{Agrawal_2017}, given by

\begin{align}
\delta_{\rm LN}(\mathbf{x}) = \exp\left[b\delta_G(\mathbf{x}) - \frac{1}{2} b^2 \sigma_G^2 \right] - 1,
\end{align}
where $\delta_G(\mathbf{x})$ is the Gaussian density contrast, $\sigma_G^2$ is its variance, and $b$ denotes the galaxy bias. The galaxy bias, $b(z)$, is calibrated by fitting theoretical angular power spectra to measurements from the MICEcat mock galaxy catalogue across multiple redshift bins \citep{Fosalba_2015}.\footnote{MICEcat adopts the same cosmology as the log-normal mocks created for this analysis.} Over the range $0.15 \leq z \leq 0.5$, the resulting best-fit quadratic model is $b(z) = 0.18z^2 + 0.52z + 1.05$. Since the gravitational wave events are generated by sampling from the galaxy distribution with a rate uniform in comoving volume, the gravitational wave and galaxy fields have the same bias functional form by construction.

These mock catalogues are used to estimate the mean measurement, as well as the covariance matrix $\mathbf{\Sigma}$ entering Equation~\ref{eq:likelihood}. Since this covariance is estimated from a finite number of realisations, $N_{\rm mocks}$, its inverse is biased. To correct for this, we apply the Hartlap correction \citep{Hartlap_2006}, which rescales the inverse covariance by a factor depending on $N_{\rm mocks}$ and the size of the data vector. In addition, the uncertainty in the estimated covariance propagates into the inferred parameter constraints. We account for this using the Percival correction \citep{Percival_2014}, which inflates the parameter covariance to capture the additional uncertainty arising from the finite number of mock realisations. Our choice of 600 mocks ensures that the Hartlap and Percival corrections have only a marginal impact on the inferred uncertainties, indicating that these effects are subdominant for the sample size considered here.

Unless otherwise stated, we adopt a luminosity distance uncertainty of $\sigma_{d_L}=1\%$, a spectroscopic-like redshift uncertainty of $\sigma_z=10^{-4}(1+z)$ \citep{Carr}, and a sky localisation in which 95\% of the probability is contained within a two-dimensional Gaussian of area 1~deg$^2$. Except where explicitly noted (see Section~\ref{sec:incompleteness}), we assume fully complete galaxy catalogues. These assumptions are optimistic relative to current gravitational wave detections. \textcolor{black}{For example, in GWTC-5.0 the median sky localisation for O4a events, observed with Virgo included in the network, was of order $100\,{\rm deg}^2$, with typical luminosity distance uncertainties of roughly $20$--$30\%$. These uncertainties depend sensitively on the detector network, signal-to-noise ratio, inclination, and source properties.} We adopt this idealised setup to isolate biases arising from modelling choices from those intrinsic to gravitational wave measurements, enabling a robust assessment of their impact on the inferred cosmological parameters.

\section{Results}
\label{sec:results}
In this section, we present the results of our analysis. We construct the data vector, $\mathbf{S}_{ij}^{\mathrm{data}}$, by measuring the angular cross-power spectra between galaxies and gravitational wave events, $C_{\ell}^{\mathrm{data}}$, using the publicly available \texttt{NaMaster} package. The corresponding theoretical $\mathbf{S}_{ij}^{\rm th}$ are computed using \texttt{CAMB}, where the galaxy and gravitational wave distributions and selection functions are implemented as source kernels and evaluated under the Limber approximation \citep{1953ApJ...117..134L}, allowing the projected clustering signal to be forward modelled in a self-consistent manner. Since each \texttt{CAMB} evaluation is computationally expensive, we compute the theoretical angular power spectra on a coarse $H_0$ grid and construct a one-dimensional cubic interpolator for $S_{ij}^{\rm th}(H_0)$.
 \begin{figure}[t]
    \centering
    \includegraphics[width=\linewidth]{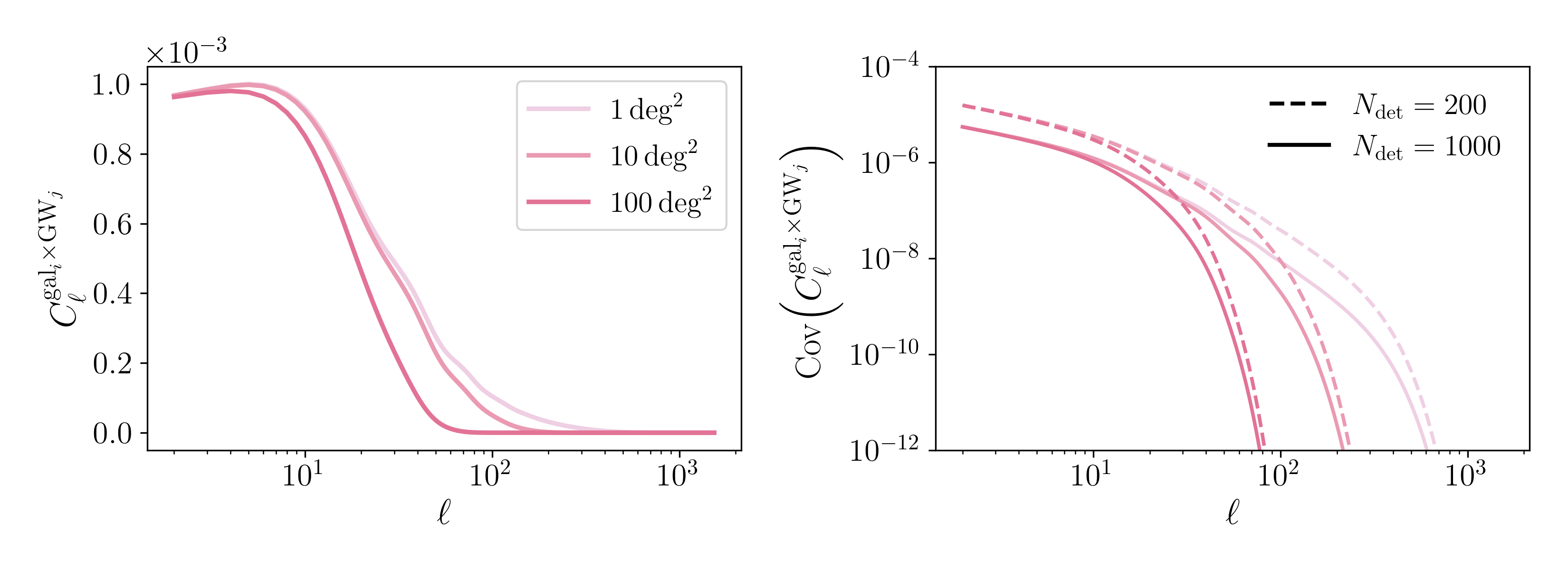}
    \caption{\textcolor{black}{The left panel shows the theoretical \texttt{CAMB} galaxy--gravitational wave cross-correlation for a single overlapping redshift and distance bin, shown for three Gaussian localisation areas. The right panel shows the corresponding diagonal element of the analytical covariance for catalogues containing $N_{\rm det}=200$ and $1000$ gravitational wave events over the full redshift range.}}
    \label{fig:clloc}
\end{figure}
Previous analyses using gravitational wave--galaxy cross-correlations model sky localisation uncertainty by smoothing the \texttt{CAMB} power spectra with a Gaussian angular kernel, thereby reproducing the smoothing observed in the data from Gaussian sky localisation areas \citep{sala2025inferringcosmologicalparametersgalaxy,pedrotti2025cosmologyangularcrosscorrelationgravitationalwave,Chen:2017rfc}. This is implemented as
\begin{equation}
C_{\ell,\mathrm{smoothed}}^{\mathrm{gal}_i\times \mathrm{GW}_j}
= W_\ell\, C_{\ell}^{\mathrm{gal}_i\times\mathrm{GW}_j},
\end{equation}
where \(W_\ell\) is the beam window function \citep{sullivan2024methodscmbmapanalysis},
\begin{equation}
W_\ell =
\exp\left[-\frac{1}{2}\,\ell(\ell+1)\,\sigma_{\mathrm{loc}}^{2}\right],
\end{equation}
with the localisation width \(\sigma_{\mathrm{loc}}\) inferred from the 95\% localisation area.\footnote{The localisation width $\sigma_{\mathrm{loc}}$ is obtained first converting to an equivalent spherical-cap radius $R$ using $A = 2\pi(1-\cos R)$, and $\sigma_{\mathrm{loc}}$ is then determined from the cumulative probability of a 2D Gaussian, $0.95 = 1-\exp[-R^2/(2\sigma_{\mathrm{loc}}^2)]$.} \textcolor{black}{As shown in Figure~\ref{fig:clloc}, poorer gravitational wave localisation areas suppress the galaxy--gravitational wave clustering signal by smoothing the gravitational wave field. This effect is strongest on small angular scales, where increasing the localisation area rapidly damps the power spectrum. The covariance is also modified, since it depends on the total observed power spectra, including the gravitational wave shot-noise term. Consequently, changing either the localisation area or the number of detections shifts the covariance, with shot noise becoming more important for smaller samples. Therefore, even when the absolute covariance is reduced in some bins, the measurement becomes less informative because the signal has been suppressed; the relevant fractional error is expected to increase for larger localisation areas. While this work adopts optimistic assumptions for the localisation areas and number of events, the implications of less ideal observational scenarios are discussed in the conclusions of each relevant section, with reference back to Figure~\ref{fig:clloc} where necessary.}

Recently, \citet{Cheng:2026atn} demonstrated that localisation uncertainty should be modelled by computing an average beam window function within each luminosity distance bin and using this to smooth the theoretical prediction. In our simplified simulations, this is exactly equivalent, as all gravitational wave events share the same localisation uncertainty.

Importantly, while we implement this smoothing in \texttt{CAMB} using an analytic beam window, knowledge of its explicit functional form is not required. In practice, the impact of angular localisation can instead be measured directly from the data by comparing the clustering of point-like event positions---where each event is treated as a delta function on the sky---with that obtained from their corresponding localisation probability maps. This provides an empirical estimate of how localisation uncertainty suppresses angular power without relying on an assumed beam model.

\subsection{Uncompressed versus compressed statistic ($C_\ell$ vs $\mathbf{S})$}
\label{sec:sijvscl}
A corresponding Gaussian likelihood can also be constructed from the full set of angular power spectra, in direct analogy with Equation~\ref{eq:likelihood} and as considered, for example, in \citet{Ferri_2025}. The advantage of using the compressed statistic is that it retains the majority of the cosmological information while significantly reducing the dimensionality of the data vector and the associated computational cost. Previous analyses (e.g. \citet{Bera_2020,Mukherjee_2024,Ferri_2025}) use the angular or 3D correlation functions in real space, of which the $S_{ij}$ can be thought of as the limiting case of purely radial separation between fields (see Appendix~\ref{app:theorysij}).

 \begin{figure}[t]
    \centering
    \includegraphics[width=\linewidth]{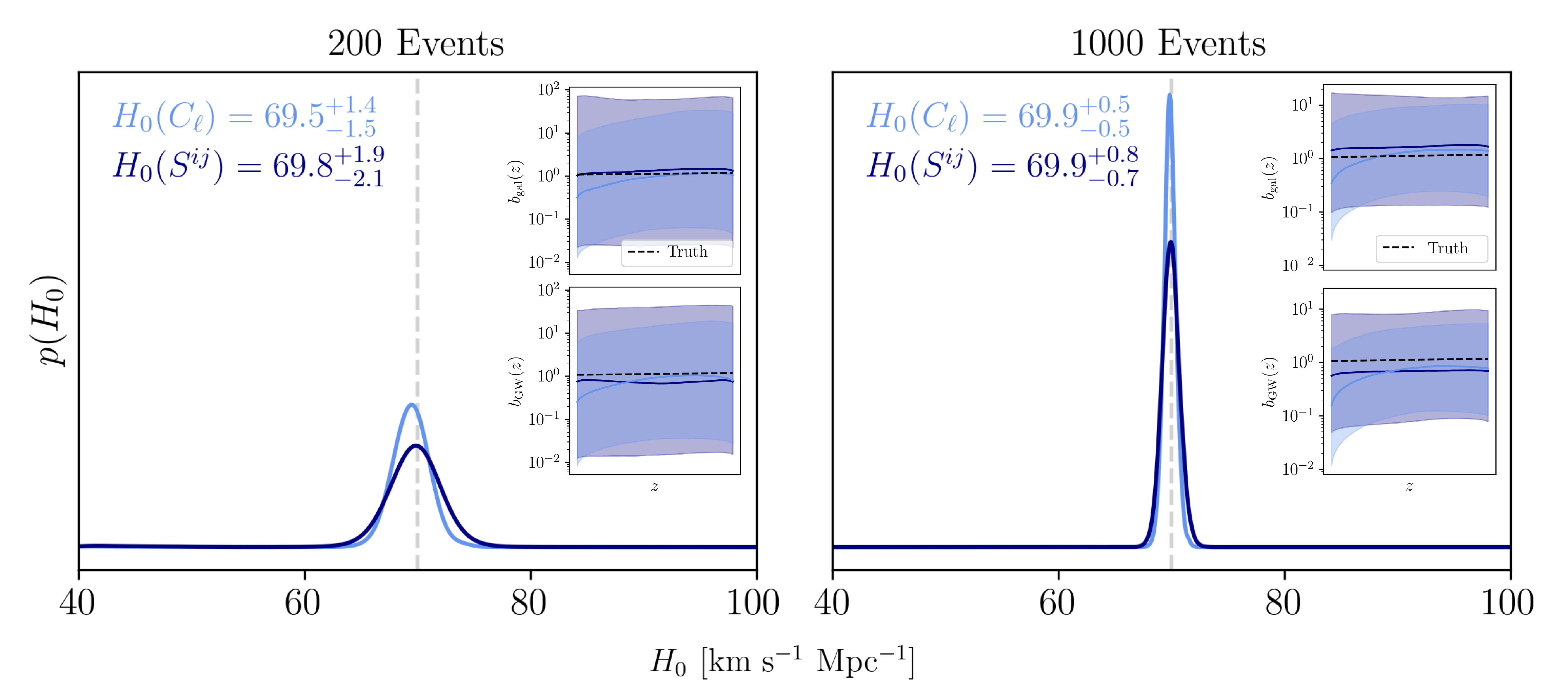}
    \caption{Comparison of the $H_0$ constraints obtained using the full cross-angular power spectra and the compressed statistic, $S_{ij}$, for 200 and 1000 events, using a diagonal analytical covariance matrix. Within each subplot, the redshift evolution of the gravitational wave and galaxy bias is shown, as inferred from the posterior samples. The black dashed line represents the underlying quadratic bias function from which the mock catalogues were generated.}
    \label{fig:sijvscl}
\end{figure}

Figure~\ref{fig:sijvscl} compares the constraints on $H_0$ obtained with the two approaches for samples of 200 and 1000 gravitational wave events, using the diagonal analytic covariance matrix in the likelihood. In both cases, the data vector is constructed from the mean over 600 mock realisations, while the quoted uncertainties correspond to the standard deviation expected from a single realisation. In all cases, we recover unbiased constraints on $H_0$. The compressed statistic leads to a modest loss of information; the uncertainty on $H_0$ increases by a factor of $\sim 1.4$ (1.5) for 200 (1000) events. This likely occurs because information from multiple scales is compressed into a single weighted quantity, thereby removing some of the constraining power carried by the individual multipoles. On the other hand, the compressed statistic offers several advantages. In addition to significantly reducing the dimensionality of the problem, it avoids the direct integration of products of Bessel functions, which makes the theoretical calculation numerically more stable without requiring the use of the Limber approximation. 

\textcolor{black}{For more realistic gravitational wave catalogues, the relative performance of the compressed statistic and the full $C_\ell$ analysis may depend weakly on the distance uncertainty and angular localisation area, although we do not expect the qualitative conclusions to change substantially. Larger distance errors broaden the radial kernel and increase leakage between neighbouring luminosity distance bins, which reduces the overall cross-correlation signal and broadens the $H_0$ constraints from both approaches. Since this primarily dilutes the signal rather than introducing new angular scale dependence, the fractional information loss from using the compressed statistic is expected to remain broadly similar. Angular localisation has a more direct effect on the multipole dependence, since as shown the right panel of Figure~\ref{fig:clloc} poorer localisation smooths the gravitational wave field on the sky and suppresses high-$\ell$ modes. This could slightly reduce the advantage of the full $C_\ell$ analysis, as there is less small-scale angular information for the uncompressed data vector to exploit.}

\subsection{Analytical versus mock covariance}
\label{subsec:cov}
Recent applications of the cross-correlation method have generally employed two techniques to quantify the uncertainty on the angular power spectrum or correlation function. The first is jackknife resampling \citep{mcintosh2016jackknifeestimationmethod} which estimates the uncertainty of individual angular cross-power spectra by repeatedly recomputing the statistic while omitting subsets of the data. While jackknife resampling captures part of the cosmic variance present within a single realisation of the data, it relies on the assumption that the jackknife regions are approximately independent. This assumption breaks down on large scales, where modes are correlated across regions, and may therefore lead to an underestimate of the true uncertainty.

The second more common approach is to compute the analytical covariance $\Sigma$ of the measured $C_\ell$ (e.g. \citet{sala2025inferringcosmologicalparametersgalaxy}, Equation~4.4). In this case, $\Sigma$ is constructed from combinations of the auto- and cross-power spectra, weighted by the survey sky fraction and multipole binning. This choice relies on the assumption that the underlying fields are Gaussian, such that the trispectrum can be neglected and the covariance expressed solely in terms of products of two-point functions. If this approximation breaks down, the resulting covariance can underestimate both the parameter uncertainties and the amplitude of off-diagonal correlations, particularly on small scales and in the presence of super-sample covariance from modes larger than the survey footprint \citep{Barreira_2018}.

We examine the impact of the covariance modelling by comparing the posteriors on the recovered $H_0$ using the compressed statistic $S_{ij}$ for 200 and 1000 gravitational wave events. Although these comparisons could, in principle, be carried out using the full angular power spectra, this would substantially increase the dimensionality of the analysis. In practice, analysing the full set of $C_\ell$ values would be infeasible for large samples of mocks due to the large number of required Hartlap--Percival corrections. We therefore restrict the covariance comparison to the compressed statistic $S_{ij}$. As shown in Figure~\ref{fig:sijvscl}, this compression produces constraints that are sufficiently consistent with those obtained from the full angular power spectra. The analytic covariance for $S_{ij}$ is obtained by propagating the analytic covariances of the individual angular power spectra $C_{\ell}$

\begin{align}
\mathrm{Cov}(S_{ij}, S_{mn}) =
\sum_\ell^{\ell_{\max}}
w_\ell^2\,
\mathrm{Cov}_\ell\!\left(
C_\ell^{\text{gal}\times \text{GW}}(i,j),\, C_\ell^{\text{gal}\times \text{GW}}(m,n)
\right),
\end{align}
where we have assumed that different angular modes are independent, i.e. only terms with $\ell = \ell'$ contribute to the covariance, with weights $w_{\ell}=\frac{2\ell+1}{4\pi}$ (which follow directly from Equation~\ref{eq:sij}). By contrast, the measured covariance is estimated directly from the 600 individual measurements of $S_{ij}$, and therefore does not require applying this analytic propagation.

 \begin{figure}[t]
    \centering
    \includegraphics[width=\linewidth]{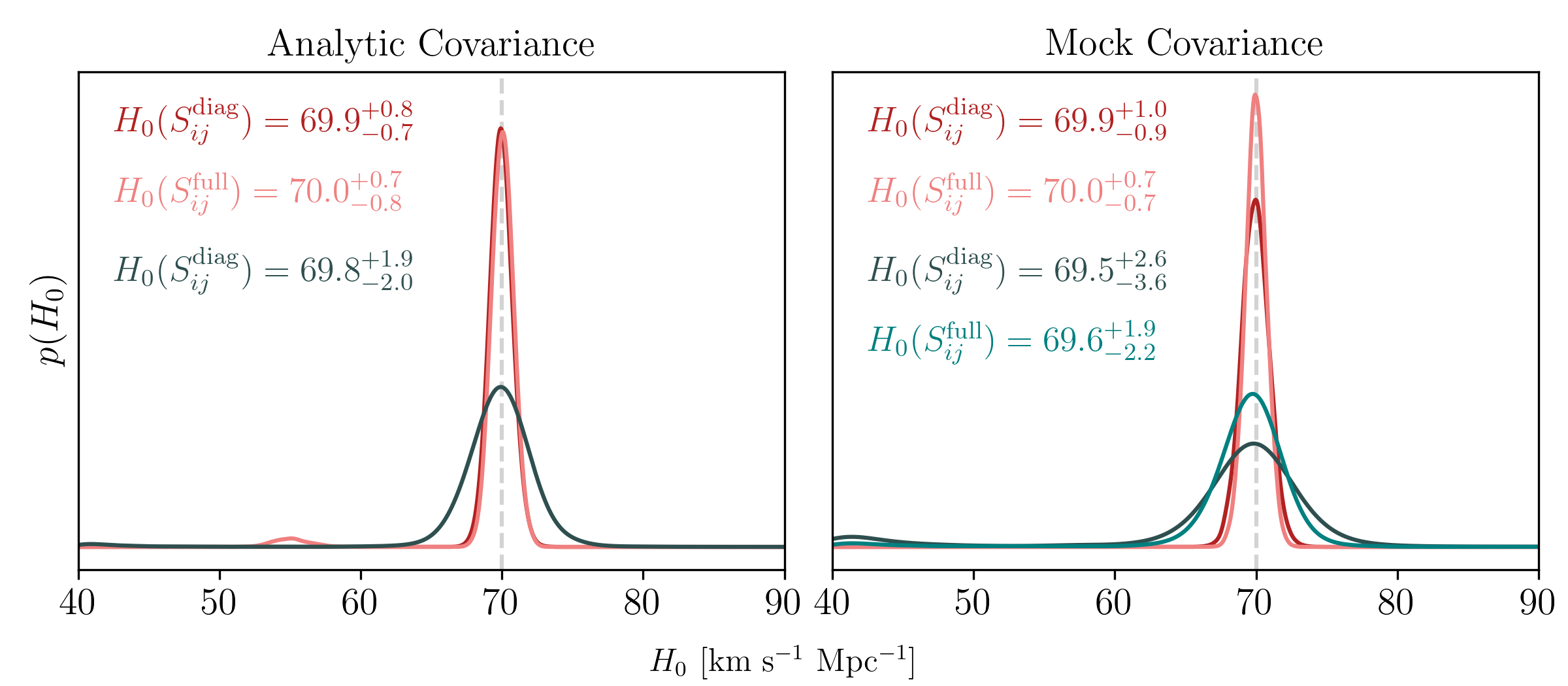}
    \caption{Comparison between the $H_0$ constraints using the analytical and mock covariance matrices (both diagonal and with the inclusion of off-diagonal elements in $(i,j)$). The green and red posteriors show the results for 200 and 1000 gravitational wave events, respectively. The $H_0$ constraints shown using the mock covariance include both the Percival and Hartlap corrections.}
    \label{fig:analytmockcov}
\end{figure}

We compare the constraints on $H_0$ obtained using both diagonal and full versions of the analytic and mock covariance matrices. In the diagonal case, we retain only covariance elements for which the two compressed statistics are identical, i.e. $\mathrm{Cov}(S_{ij},S_{mn})$ is kept only when $i=m$ and $j=n$. In the full case, we include all covariance elements between different redshift and luminosity distance bins, allowing correlations between $S_{ij}$ and $S_{mn}$ for arbitrary combinations of $(i,j)$ and $(m,n)$. The results are presented in Figure~\ref{fig:analytmockcov}. As before, the data vector is constructed from the mean over 600 mock realisations, while the quoted uncertainties correspond to the standard deviation across mocks, representing the uncertainty expected for a single realisation. For $10^3$ events, we find that the analytical and the mock covariances perform equally well when the off-diagonal terms are included in the analysis. We also notice that including the off-diagonal covariance elements leads to a reduction in the inferred uncertainty on $H_0$ when using the mock covariance, indicating the presence of non-negligible correlations between bins. Using the off-diagonal terms in the analytical covariance, on the other hand, does not change the magnitude of the uncertainties, indicating that the analytical covariance formula might underestimate the impact of correlations between bins.

For 200 events in our fiducial sky distribution, the diagonal analytical covariance is found to underestimate the uncertainty on $H_0$ compared to the mock diagonal covariance by a factor of $\times 1.5$. Furthermore, the inclusion of non-diagonal elements in the shot-noise-dominated regime makes the inversion of the analytical covariance matrix numerically unstable, making it impossible for the MCMC parameter exploration to converge and produce meaningful constraints on $H_0$.\footnote{Hence why there is no $S_{ij}^{\rm full}$ in the left panel of Figure~\ref{fig:analytmockcov} for 200 events.} The mock off-diagonal covariance yields tighter constraints than the only-diagonal one, consistent with the case of 1000 events, though the difference between the diagonal and full covariance is found to be larger for fewer events. 

\textcolor{black}{For less idealised gravitational wave catalogues, the comparison between the mock and analytic covariance estimates may become more important. Larger distance uncertainties increase leakage between neighbouring luminosity distance bins, which can enhance correlations between different $S_{ij}$ measurements and may not be fully captured by a simplified analytic covariance. This suggests that the mock covariance could provide a more reliable estimate of the off-diagonal covariance structure in this regime (although with the caveat of requiring more mocks). The high-$\ell$ mode suppression from poorer angular localisation areas may reduce the importance of small-scale structure and make the analytic Gaussian covariance approximation more accurate. We therefore expect the analytic covariance to remain useful for capturing the broad uncertainty scaling, but for more realistic catalogues the mock covariance may be needed to robustly quantify inter-bin correlations and the effect of distance-bin leakage.}

Overall, this comparison highlights the importance of using covariance estimates derived from mocks, as analytical approaches may yield overly optimistic constraints, particularly in the case of smaller numbers of gravitational wave events. It also demonstrates that off-diagonal correlations play a non-negligible role and can significantly affect the final parameter uncertainties.

\subsection{Systematic uncertainties}
\label{sec:systematics}
Having verified that the use of the compressed statistics $S_{ij}$ with the appropriate choice of covariance recovers the correct fiducial cosmology, we are now in the position of investigating the main sources of systematic uncertainty affecting the cross-correlation method, namely distance uncertainties, galaxy and gravitational wave bias mis-parametrisation, and incompleteness.

\subsubsection{Malmquist and Eddington bias}
\label{subsec:malmqedd}
Luminosity distance uncertainty was identified in \citet{crossparkin2025darksirensimpactredshift} as a primary limitation in obtaining competitive constraints on $H_0$ in the galaxy catalogue approach. The cross-correlation method is significantly less sensitive to luminosity distance uncertainty than the galaxy catalogue approach, provided that the binning is chosen such that the characteristic width of the luminosity distance bins exceeds the typical distance uncertainty of the gravitational wave events. In this regime, the impact of distance uncertainty is reduced, as events are less likely to be scattered between bins. The choice of binning will be discussed further in Section~\ref{subsec:binning}.

\begin{figure}[t]
    \centering
    \includegraphics[width=\linewidth]{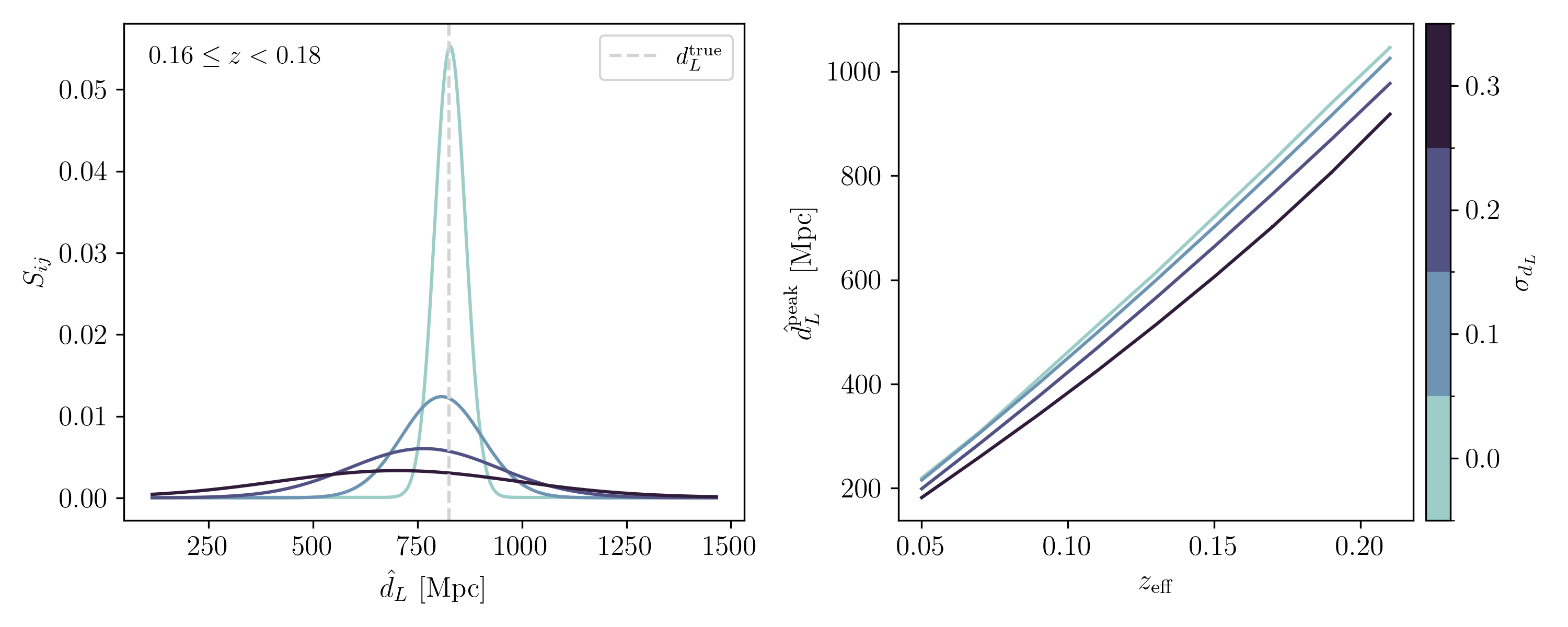}
    \caption{Illustration of how luminosity distance uncertainty affects the inferred distance--redshift relation. The left panel shows the compressed cross-correlation signal, $S_{ij}$, as a function of observed luminosity distance, $\hat{d}_L$, for the redshift bin $0.16 \leq z < 0.18$, with colours indicating the fractional distance uncertainty, $\sigma_{d_L}$. The vertical dashed line indicates the true luminosity distance corresponding to the bin’s effective redshift. The right panel shows the resulting peak distance, $\hat{d}_L^{\rm peak}$, as a function of effective redshift, $z_{\rm eff}$, for the same values of $\sigma_{d_L}$.}
    \label{fig:malmqedd}
\end{figure}

However, large uncertainties in the inferred luminosity distances of gravitational wave sources do introduce both Eddington bias \citep{1913MNRAS..73..359E} and Malmquist bias \citep{1920MeLuS..22....3M}, which can significantly affect cross-correlation measurements of $H_0$ if not correctly modelled. 
These biases enter the analysis at the point where the true luminosity distance of each gravitational wave source is converted into an observed luminosity distance and then subjected to a detection threshold. The observed distance is obtained by scattering the true distance with a fractional uncertainty in linear space (i.e. $\hat{d}_L$ is drawn from a distribution $\mathcal{N}(d_L^{\rm true}, \sigma_{d_L}d_L^{\rm true}$). This measurement scatter introduces Eddington bias: because the underlying source population rises with distance, more events are scattered from larger true distances into smaller observed distances than the reverse, shifting the observed distance distribution toward lower $d_L$. A further distortion is introduced by Malmquist bias, which arises from the detection threshold: events scattered to larger observed distances are more likely to fall below the signal-to-noise cut and be lost, whereas events scattered inward remain detectable. Together, as shown in Figure~\ref{fig:malmqedd}, Eddington and Malmquist biases preferentially shift the observed gravitational wave population toward smaller inferred luminosity distances by enhancing downward scatter while suppressing events scattered to larger distance, thereby biasing the recovered $H_0$ high. Interestingly, as illustrated by the Hubble diagram in the right panel of Figure~\ref{fig:malmqedd}, these biases influence not only the inferred value of $H_0$, but also the overall shape of the Hubble diagram, and hence the constraints on $\Omega_m$.

Eddington and Malmquist biases are incorporated in the theoretical prediction through forward modelling of the gravitational wave selection function. Instead of assigning sources at redshift $z$ to a luminosity distance bin using the true $d_L$, the theory computes the probability that a source is observed in a given bin after accounting for measurement uncertainty and selection effects (given by Equation~\ref{eq:probdl}). In particular, the use of the cumulative distribution function $\Phi$ mitigates the Eddington bias, and the introduction of a step function suppression above the threshold in the probability $P(\hat{d}_L| d_L^{\mathrm{thr}})$ mitigates Malmquist bias. Implementing these corrections into the gravitational wave kernel entering the angular power spectrum (Equation~\ref{eq:kernel2simp}) ensures that the effects of distance scatter and detection thresholds are propagated consistently into the theoretical prediction. As a result, the theoretical model naturally reproduces the same biases present in the observed data.

\textcolor{black}{This correction is particularly important for current gravitational wave events, for which distance uncertainties are large and scattering between distance bins can lead to significant biases in the inferred parameters if not properly modelled. For future detections with substantially smaller distance errors, this bin-to-bin scattering will be reduced, and the corresponding Eddington and Malmquist corrections are therefore expected to decrease. However, so to will the statistical errors. So it will remain important to compare the magnitudes of the statistical and systematic errors for future analyses with greater numbers of detections or more precise gravitational wave measurements.}

\subsubsection{Incorrect bias parametrisation}
\label{subsec:biasparam}
Cosmological parameter inference using the cross-correlation method requires the bias parameters of both galaxies and gravitational waves to be treated as nuisance parameters and jointly fitted. Much of the current literature models these as linear $b_x(1+z)$ across the sample (e.g. \citet{Cheng:2026atn,Ferri_2025}) or constant in each tomographic bin (e.g. \citet{pedrotti2025cosmologyangularcrosscorrelationgravitationalwave}). These modelling choices, especially when relatively large bins are considered, might introduce biases in the inferred cosmological parameters. 

Since the true functional form of the bias is not known \textit{a priori}, it becomes important to explore the impact of mis-parametrising the functional form of the biases. As a test case, we study how the constraints on the cosmological parameters change adopting different polynomial parametrisations up to quadratic order (where the truth has quadratic order). We allow the gravitational wave and galaxy samples to have separate bias parameters, rather than enforcing them to be identical, even though they are the same in the simulations. This choice reflects the more realistic case of applying the method to data, where the two bias functions would not be known in advance.

 \begin{figure}[t]
    \centering
    \includegraphics[width=\linewidth]{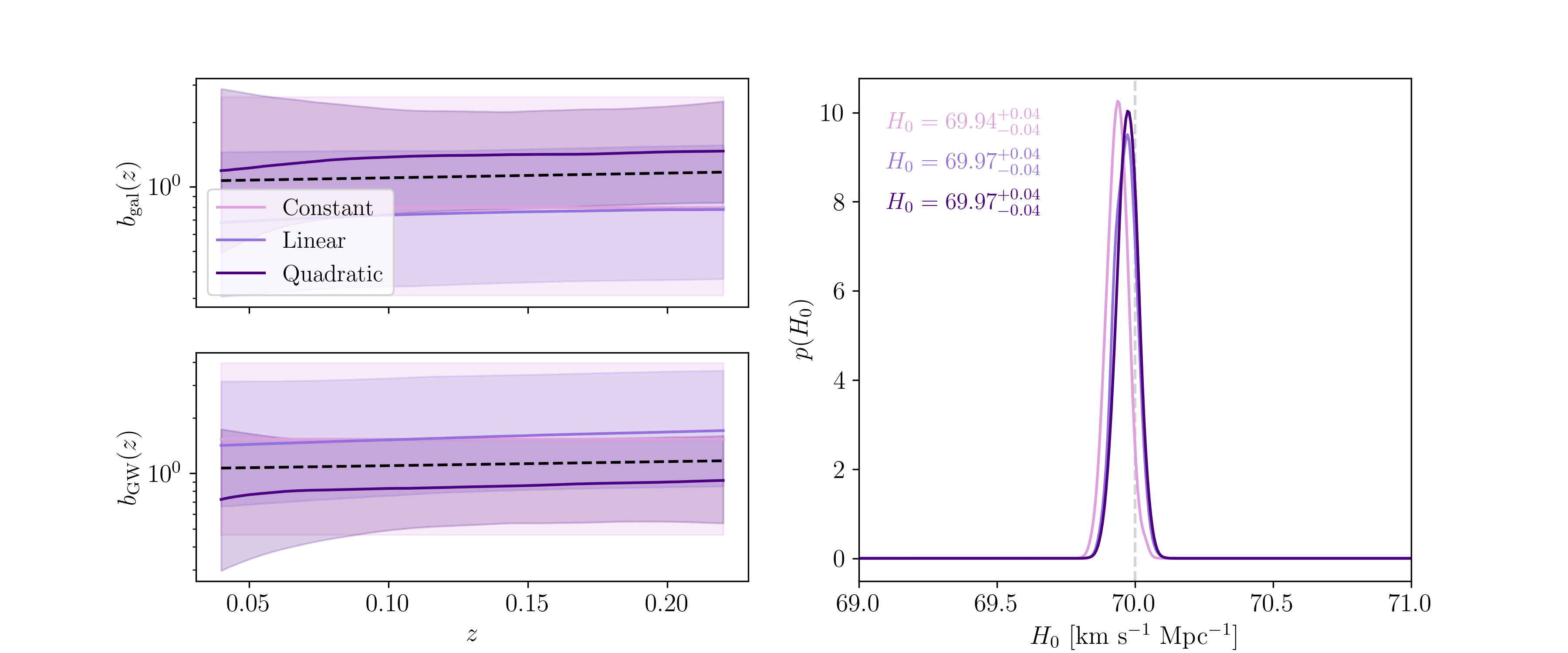}
    \caption{Resulting constraints on the bias parameters and $H_0$ for constant, linear, and quadratic parametrisations of the galaxy and gravitational wave bias. The black dashed line indicates the true underlying quadratic bias. The $H_0$ constraints shown here are obtained using the mock diagonal covariance matrix, including the the Hartlap and Percival corrections.}
    \label{fig:bgwbgal}
\end{figure}

The results of varying the parametrisation of the two biases in the likelihood of Equation~\ref{eq:likelihood}, using 1000 gravitational wave events, are shown in Figure~\ref{fig:bgwbgal}. Here, the mock mean and the standard error on the mean are used to isolate systematic shifts in $H_0$ from the intrinsic scatter between realisations. We see that the linear and quadratic bias are able to obtain unbiased $H_0$ estimates, whereas the constant parametrisation does not appear to have the flexibility to fully capture the redshift evolution of the true bias, leading to a biased $H_0$ measurement. \textcolor{black}{However, this bias is expected to be subdominant for current and near-future gravitational wave observations, once realistic distance uncertainties and localisation areas are included, since these effects substantially increase the statistical uncertainty.} This implies that accurate cosmological constraints can be obtained without precise knowledge of the bias evolution, although an insufficiently flexible parametrisation can still introduce systematic biases if it fails to adequately describe the underlying behaviour.

\subsubsection{Biases due to incompleteness}
\label{sec:incompleteness}
One of the advantages of the cross-correlation method over the standard galaxy catalogue approach is its reduced sensitivity to incompleteness due to selection effects \citep{Bera_2020}. To illustrate why this is the case, consider a simple galaxy detection probability $p_{\rm det}(z)$. If the true galaxy number density of the underlying (unlimited) sample is
\begin{equation}
    n(\mathbf{x}, z) = \bar n(z)\,[1 + \delta(\mathbf{x}, z)],
\end{equation}
then a detection probability that depends only on redshift---such as a magnitude limit---acts as a position-independent selection function, uniformly subsampling the galaxy population at fixed $z$. This rescales the mean number density while leaving spatial fluctuations unchanged, such that the detected density can be written as
\begin{equation}
\label{eq:ndet}
    n_{\rm det}(\mathbf{x}, z)
    = p_{\rm det}(z)\,\bar n(z)\,[1 + \delta(\mathbf{x}, z)].
\end{equation}
The corresponding overdensity is therefore
\begin{align}
\delta_{\mathrm{det}}(\mathbf{x}, z)
&=
\frac{n_{\mathrm{det}}(\mathbf{x}, z) - \bar n_{\mathrm{det}}(z)}
{\bar n_{\mathrm{det}}(z)} \\
&=
\frac{p_{\mathrm{det}}(z)\,\bar n(z)\,[1 + \delta(\mathbf{x}, z)] - p_{\mathrm{det}}(z)\,\bar n(z)}
{p_{\mathrm{det}}(z)\,\bar n(z)} \\
&= \delta(\mathbf{x}, z).
\end{align}

This cancellation holds only if the selection function, $p_{\rm det}(z)$, is known and consistently included in the expected mean density, such that $\bar n_{\rm det}(z)=p_{\rm det}(z)\bar n(z)$. In this case, a purely radial selection function simply rescales the mean number density and leaves the overdensity field unchanged, provided that each bin contains enough galaxies for the measured tracer overdensity to faithfully trace the underlying density field. However, such a selection can modify the effective galaxy bias as brighter galaxies are typically more strongly biased. While this changes the amplitude of the cross-correlation signal, fitting the bias parameters simultaneously with $H_0$ absorbs this rescaling, ensuring that the inferred value of $H_0$ remains unbiased. 

To illustrate this, we consider three samples defined by light, hard, and combined selection criteria. In MICEcat, these correspond to apparent magnitude cuts. Since the log-normal mocks do not include galaxy magnitudes, we implement the same effects through equivalent radial selection functions:

\begin{itemize}
    \item [i.] $p_{\rm light}(z)$: radial selection corresponding to an apparent magnitude cut of $m_r\leq 21$ in MICEcat.
    \item [ii.] $p_{\rm hard}(z)$: radial selection corresponding to an apparent magnitude cut of $m_r\leq 19$ in MICEcat.
    \item [iii.] $p_{\rm light + hard}(z)$: radial selection corresponding to a declination-dependent apparent magnitude cut of $m_r\leq 19$ for $0^\circ\leq \delta\leq 45^\circ$ and  $m_r\leq 21$ for $45^\circ\leq \delta\leq 90^\circ$ in MICEcat.
\end{itemize}

The third selection would be analogous to combining regions of differing completeness such as those observed in surveys like DESI \citep{DESI_Collaboration_2022} and 4MOST \citep{2012SPIE.8446E..0TD}. \textcolor{black}{While the declination cut used here is a simplified representation of such survey selection effects, it provides a useful test case for demonstrating how a known incompleteness can be consistently forward-modelled. More realistic effects, such as fibre collisions, variations in observing conditions, and target-dependent redshift success rates, are routinely modelled in large-scale-structure surveys. These effects could therefore be imprinted on a random catalogue and incorporated into the selection function entering Equation~\ref{eq:ndet}, thereby contributing to the effective $p_{\rm det}$.}

\begin{figure}[t]
    \centering
    \includegraphics[width=\linewidth]{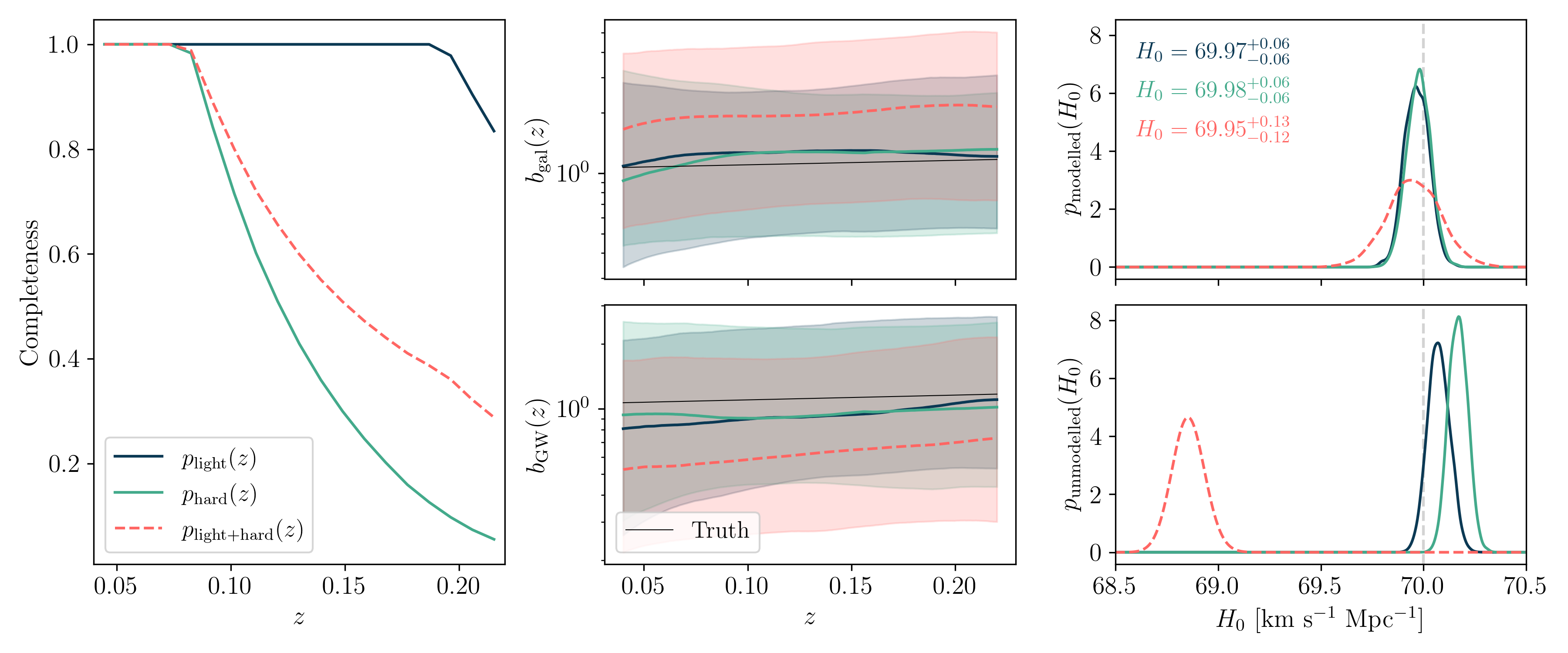}
    \caption{Results for three radial selection functions using the diagonal elements of the mock covariance matrix. \textit{Left:} completeness fraction as a function of redshift for each sample. \textit{Middle:} constraints on the gravitational wave and galaxy bias parameters for the three samples. \textit{Right:} constraints on $H_0$ for the three samples, comparing cases where the selection function is forward modelled (top) and not included in the theory (bottom).}
    \label{fig:compl}
\end{figure}

The results for these three sample's cuts are shown in Figure~\ref{fig:compl}. In each case, we use the mean over mock realisations and report the standard error on the mean, with each scenario generated using 1000 events. From the top-right panel of Figure~\ref{fig:compl}, we see that unbiased constraints on $H_0$ can be recovered in the presence of radial selection effects, provided that the selection function is consistently forward modelled in the theoretical prediction entering the likelihood in Equation~\ref{eq:likelihood}. This is implemented by multiplying the gravitational wave kernel by the completeness selection function (given by the left panel in Figure~\ref{fig:compl}), ensuring that the theoretical prediction reflects the same radial weighting as the observed data. In practice, this selection function may not be explicitly known, so an alternative approach is to use the observed redshift distribution to construct the effective radial window function, allowing the selection effects to be captured empirically rather than through an assumed model. The downside is that the observed radial distribution is itself subject to statistical uncertainties, which can introduce additional noise into the theoretical model and weaken the inferred constraints on $H_0$.

Although the covariance is larger in the $p_{\rm hard}(z)$ sample due to increased shot noise from fewer galaxies, this does not lead to a corresponding increase in the uncertainty on $H_0$. This highlights that an increase in covariance on certain scales cannot be translated directly into a change in the $H_0$ uncertainty. The resulting uncertainty is set by the interplay between cosmic variance, galaxy/gravitational wave shot noise, and degeneracies between $H_0$ and the bias parameters, which can lead to non-intuitive behaviour. In particular, changes in the selection function can redistribute information across parameter space, such that some loss of information may occur in directions orthogonal to $H_0$ and therefore have little impact on its marginalised uncertainty.

For the combined $p_{\rm light+hard}(z)$ case, we do see an increase in uncertainty. This behaviour arises from how the bias parameters are constrained. The shallower catalogue lacks galaxies at higher redshift, while the deeper catalogue extends to larger distances. When the two are combined, the MCMC must fit the bias parameters independently in each redshift bin over this broader range, including regions where one of the samples provides little constraining power. This leads to a degradation in the overall constraints. As a result, the combined sample effectively inherits the wider redshift sensitivity of the deeper catalogue, but with reduced constraining power in parts of that range, leading to larger overall uncertainties.

While unbiased constraints on $H_0$ are recovered in all three scenarios, this is only achieved when the radial selection is consistently incorporated into the theoretical model. In the bottom-right panel of Figure~\ref{fig:compl}, the theory is instead fixed to that of the full catalogue. In this case, the likelihood compares data and theory with inconsistent radial weights, leading to a biased inference of $H_0$. This effect is particularly pronounced in the combined $p_{\rm light+hard}(z)$ case, where the mismatch between the true and assumed radial weighting is more complex. However, the results shown here are based on the standard error on the mean; for a single realisation, the uncertainties would be significantly larger, and the resulting bias is therefore expected to be negligible. \textcolor{black}{Moreover, the bias seen when the selection is not forward-modelled is modest even in this highly optimistic case. For current and near-future gravitational-wave samples, where statistical uncertainties are much larger, this bias is likely to remain subdominant. }

Forward modelling the selection function in the theory highlights a fundamental distinction from the standard dark siren approach. In that framework, the observed redshift distribution cannot be used directly in the likelihood, and one must instead model the full underlying galaxy population, including survey incompleteness and the contribution from unobserved galaxies. By contrast, the cross-correlation method depends only on the observed overdensity field, with selection effects entering solely through the radial weighting, allowing the empirically measured galaxy distribution to be incorporated directly into the theoretical prediction and making the method robust to complex or poorly characterised selection effects without requiring detailed modelling of the missing population.

\subsection{Impact of binning}
\label{subsec:binning}
The choice of redshift and luminosity distance binning in a tomographic analysis can significantly impact the inferred cosmological parameters. Its constraining power is governed by a trade-off between shot noise and signal dilution: finer bins provide greater resolution of structure—effectively increasing the number of data points—but suffer from higher shot noise due to fewer objects per bin. Conversely, broader bins reduce shot noise at the expense of washing out redshift and distance information.

\begin{figure}[t]
    \centering
    \includegraphics[width=\linewidth]{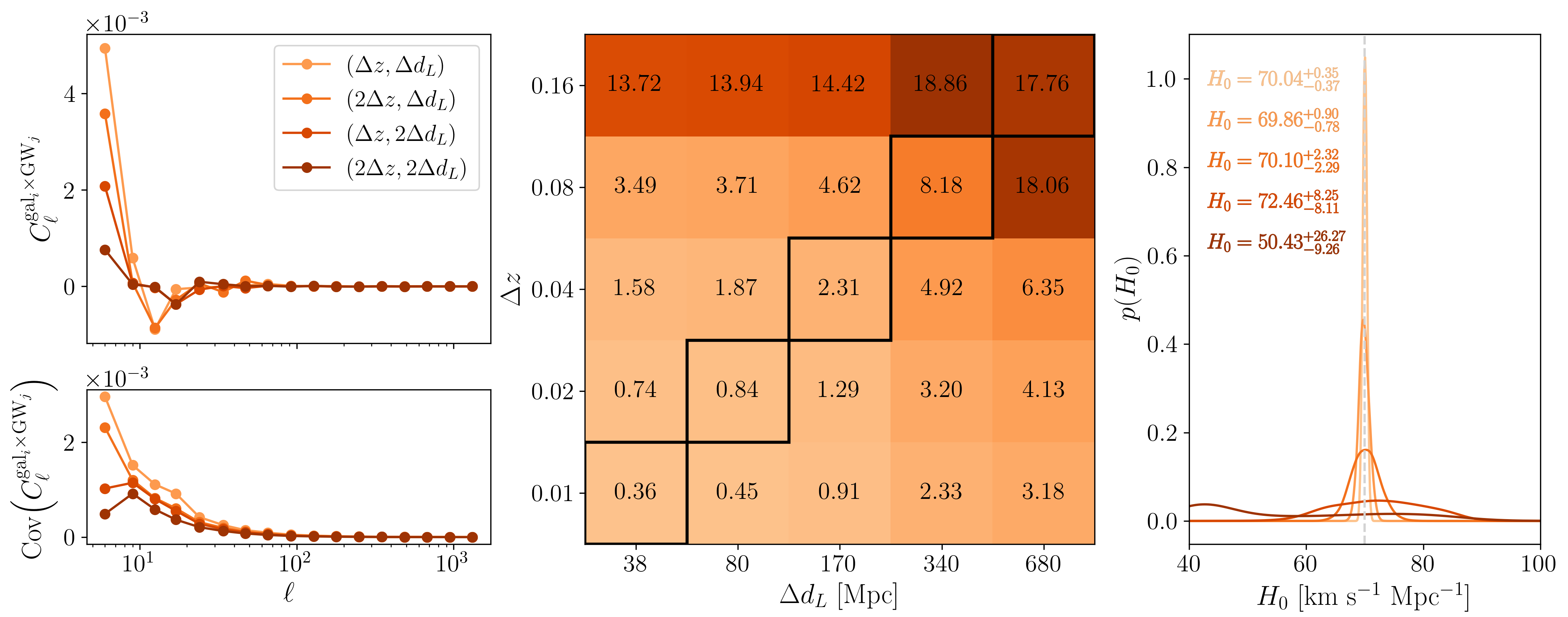}
    \caption{The left panel shows the angular power spectra and corresponding covariance for four choices of redshift and distance binning in an overlapping-bin configuration. Here, $(\Delta z,\Delta d_L)$ denotes the fiducial binning, while factors of 2 indicate bins that are twice as wide in the corresponding dimension. The middle panel shows how the choice of tomographic binning impacts the $H_0$ constraint, with the numbers in each box indicating the corresponding uncertainty in $H_0$, for both independent changes in redshift and luminosity distance binning and cases where both are varied together. The right panel shows the corresponding $H_0$ posteriors for the diagonal configurations in the middle panel. All of these results are computed using the diagonal elements of the analytical covariance matrix.}
    \label{fig:binning}
\end{figure}

Our fiducial binning scheme is chosen to balance these competing effects, retaining sufficient tomographic resolution while maintaining adequate signal-to-noise in each bin. Figure~\ref{fig:binning} illustrates how varying the luminosity distance and redshift binning, both jointly and independently, affects the constraint on $H_0$ for 1000 events. In this case, we use 100 mocks and quote the mean and standard deviation expected for a single realisation.

It is clear that decreasing the number of redshift bins has the largest detrimental impact on the uncertainty on $H_0$. This behaviour is primarily driven by the differing role of shot noise in the two dimensions. Galaxies provide a dense sampling of the underlying matter field, so even with fine redshift binning the associated shot noise remains small, allowing the additional radial resolution to directly improve the signal. In contrast, gravitational wave events are comparatively sparse, and subdividing the luminosity distance bins significantly increases the shot noise. Consequently, finer redshift binning leads to a genuine gain in constraining power, while finer luminosity distance binning yields diminishing returns as the measurements become increasingly noise dominated.

An important consideration arises when accounting for uncertainties in the gravitational wave distances and galaxy redshifts. In the traditional galaxy catalogue approach, the luminosity distance uncertainty has a particularly strong impact on the uncertainty in $H_0$. This is because the inference relies on matching individual gravitational wave events to host galaxies, such that uncertainty in the luminosity distance directly broadens the mapping between distance and redshift, degrading the precision of the measurement.

In contrast, the cross-correlation method is less sensitive to this effect. As long as the bin widths are comparable to or larger than the typical luminosity distance uncertainty, the impact of this uncertainty is effectively absorbed into the binning. Rather than directly degrading the constraint on $H_0$, the primary effect is instead mediated through the choice of bin width, with larger bins reducing the overall sensitivity (as shown in Figure~\ref{fig:binning}). \textcolor{black}{For current gravitational wave catalogues, the large distance uncertainties and limited number of detections favour the use of relatively broad distance bins. Large distance errors scatter events between neighbouring bins, while small event numbers increase the shot noise in each bin. Provided this bin leakage is consistently included in the likelihood, as discussed in Section~\ref{subsec:malmqedd}, it should not bias the inferred value of $H_0$. Its main effect is instead to wash out the cross-correlation signal, leading to broader constraints. The conclusions about the optimal redshift and luminosity distance binning are not expected to depend strongly on the angular localisation area. As shown in Figure~\ref{fig:clloc}, angular localisation primarily suppresses high-$\ell$ modes, whereas the binning choice mainly controls the radial resolution; poorer localisations may simply reduce the need for very fine angular-scale resolution.} The exact interplay between measurement uncertainties and binning-induced effects must be carefully assessed, as the optimal balance will likely vary on a case-by-case basis.

This trade-off can be quantified by considering the Fisher information for $H_0$
\begin{align}
\label{eq:fisher}
    F=\sum_{i,j,m,n}\left(\frac{\partial S_{ij}}{\partial H_0}\right) \left[\mathrm{Cov}\left(S_{ij},S_{mn}\right)\right]^{-1}\left(\frac{\partial S_{mn}}{\partial H_0}\right),
\end{align}
where the corresponding statistical uncertainty on $H_0$ is
\begin{align}
    \sigma(H_0) = \frac{1}{\sqrt{F}} \,.
\end{align}

Qualitatively, three intertwined effects determine how the Fisher information changes with bin size. First, the signal $S_{ij}$ scales inversely with the bin width (see Equation~\ref{eq:binning_change}), and since the binning itself is independent of $H_0$, the same scaling applies to $\partial S_{ij}/\partial H_0$. Second, the uncertainty on $S_{ij}$ also increases as the bins become smaller, reflecting the reduced number of sources per bin. Finally, decreasing the bin size increases the number of terms in the sum in Equation~\ref{eq:fisher}, with off-diagonal elements of the covariance contributing non-trivially due to correlations between neighbouring bins.

The first two effects are illustrated in the left panel of Figure~\ref{fig:binning}, which shows the signal (top) and noise (bottom) for a reference $C_\ell$ between two overlapping bins. The third effect is less straightforward to interpret qualitatively; however, the middle panel of Figure~\ref{fig:binning} shows that smaller bins consistently yield tighter constraints on $H_0$. This suggests that the additional covariance contributions either track the scaling of the signal or are not sufficiently large to counteract it.

The behaviour observed in Figure~\ref{fig:binning} can be understood quantitatively by considering the following approximation for $C_\ell^{ij}$,
\begin{align} \label{eq:binning_change}
\begin{split}
    C_\ell^{ij}&=\int_0^\infty dk\int_{z^{\mathrm{min}}_i}^{z^{\mathrm{min}}_i+\Delta z_i} dz_1 \int_{z^{\mathrm{min}}_j}^{z^{\mathrm{min}}_j+\Delta z_j} dz_2 \; k^2 P_m(k,z_1,z_2) W^i(z_1)W^j(z_2)j_{\ell}(k\chi_1) j_\ell(k\chi_2)\;\\
    &\approx \int\int dz_1\; dz_2\; \frac{\delta_D\left(\chi_1 - \chi_2\right)}{\chi^2} \;P_m\left(\frac{\ell + \frac{1}{2}}{\chi},z_1,z_2\right) W^i(z_1)W^j(z_2)\;\\
    &\approx \frac{\Delta z_{\mathrm{overlap}}}{\Delta z_i \Delta z_j}\frac{1}{(\chi_{\mathrm{min}}^{\mathrm{overlap}})^2}P_m\left(\frac{\ell + \frac{1}{2}}{\chi_{\mathrm{min}}^{\mathrm{overlap}}},z_{\mathrm{min}}^{\mathrm{overlap}}\right)\;,
    \end{split}
\end{align}
where the second line follows from the Limber approximation and the final expression is obtained by expanding around the bin widths $\Delta z_i$ and $\Delta z_j$, as well as their overlapping region $z_{\mathrm{min}}^{\mathrm{overlap}} < z < z_{\mathrm{min}}^{\mathrm{overlap}} + \Delta z_{\mathrm{overlap}}$. The explicit dependence on $\Delta z_i$ and $\Delta z_j$ arises from the normalisation of the window functions, which introduces the inverse scaling with bin width
\begin{equation}
    W^{i,j}(z_{i,j})\approx (\Delta z_{i,j})^{-1}\;.
\end{equation}
For perfectly overlapping bins, where the signal is maximised, it follows that $C_\ell^{ij} \propto 1/\Delta z_i$. Similarly, the shot noise in each bin scales as $N^i \propto (\Delta z_i)^{-1}$, reflecting the inverse dependence on the number of sources per bin. This immediately explains the behaviour observed in Figure~\ref{fig:binning}, in which both the signal and noise scale inversely with bin size. In particular, the $\sim 4$ increase in the covariance when halving the bin width indicates that the shot noise in the gravitational wave distribution dominates the covariance matrix, which therefore scales as $\propto (N^{\rm GW})^2$. Finally, the structure of Equation~\ref{eq:binning_change} shows that the $H_0$-dependent terms---entering through the $\chi^2$ factor in the denominator and the shape of the matter power spectrum $P_m(k)$---separate from the explicit bin-width dependence. This implies that variations in $H_0$ and $\Delta z$ effectively commute, validating the analytical reasoning behind the impact of binning choices on the Fisher information (that is, an increase of signal $S_{ij}$ is associated to an increase in $F$) in Equation~\ref{eq:fisher}.

\section{Discussion}
\label{sec:discussion}

Cross-correlation analyses have been recently suggested as an alternative to the standard galaxy catalogue approach for inferring cosmological parameters using gravitational wave events. In this work, we complemented the existing literature by performing an in-depth exploration and comparison of different methodological choices and systematic biases common in large-scale structure analyses. 
In particular:
\begin{itemize}
    \item Most of the existing literature on the angular cross-correlation between gravitational waves and galaxies use either the real space angular correlation function (e.g. \citet{Mukherjee_2024,Bera_2020}) or the angular power spectra (e.g. \citet{Oguri_2016,Cheng:2026atn}). The dimensionality of the problem, especially for analyses involving a large number of bins, can therefore become computationally expensive very quickly. Furthermore, if the Limber approximation is not used, the oscillatory behaviour of the spherical Bessel functions appearing in the integrals for the $C_\ell$ can introduce numerical instabilities. These issues can be mitigated by compressing the angular power spectra information into a single summary statistic $S_{ij}$. As we showed in Section~\ref{sec:sijvscl}, the latter largely preserves the cosmological information (see Figure~\ref{fig:sijvscl}). Appendix $\ref{app:theorysij}$ discusses the simple physical interpretation of $S_{ij}$ as the line-of-sight cross-correlation of the two tracers, i.e. the real space correlation function along the radial separation.
    
    \item We compared in Section~\ref{subsec:cov} the impact of using an analytical covariance matrix versus one computed from 600 log-normal mock realisations. Our results, reported in Figure~\ref{fig:analytmockcov}, show that the two approaches are equivalent if a sufficiently large number of events is considered ($\sim 10^3$) and the off-diagonal terms are properly taken into account. In particular, we find that including the off-diagonal components in the mock-derived covariance leads to a reduction in the uncertainty on $H_0$. For fewer event numbers, when the full covariance is modelled analytically, the off-diagonal correlations between bins make the inversion of the covariance matrix numerically unstable, which prevents the convergence of the MCMC parameter exploration and the recovery of meaningful constraints. Furthermore, using only the diagonal part of the covariance matrix led to a slight underestimation of the uncertainties compared to the full non-diagonal mock covariance. Combined with the ability of mocks to incorporate realistic survey geometries, this underscores the importance of using mock based covariance estimates.

    \item In Section~\ref{subsec:malmqedd} we show explicitly that Eddington and Malmquist biases, if not properly accounted for, lead to a systematic underestimation of the true distance. As shown in Figure~\ref{fig:malmqedd}, this is particularly important for catalogues with large distance uncertainties $\sigma_{d_L}$. In our analysis, these effects can be straightforwardly accounted for by forward modelling the detection probability of an observed gravitational wave distance using a combination of the cumulative distribution function $\Phi$ and an appropriate detection threshold $d_L^{\rm thr}$, as described in Equations~\ref{eq:probdl} and~\ref{eq:probz}.

    \item We showed explicitly in Section~\ref{subsec:biasparam} that the redshift evolution of the gravitational wave and galaxy biases can be forward-modelled in our theory by defining appropriate weighted kernel averages of the bias product (see Equation~\ref{eq:Aij}) for each pair of distance and redshift shells. Although this effect is expected to be negligible for near-future observations, Figure~\ref{fig:bgwbgal} shows that if the bias parametrisation is not sufficiently flexible, the resulting posterior on $H_0$ can be systematically biased.
    
    \item In Section~\ref{sec:incompleteness} we confirm explicitly that incomplete catalogues can still yield unbiased constraints on $H_0$, provided that selection effects are consistently forward modelled in the theoretical prediction. This represents an advantage over the standard galaxy catalogue approach, as incompleteness is naturally incorporated into the radial window function, rather than requiring explicit repopulation or modelling of the missing galaxies. On the other hand, we show in Figure~\ref{fig:compl} that combining catalogues with different completeness in different regions of the sky can significantly increase the uncertainties in the recovered cosmological parameters. This occurs because combining different galaxy catalogues significantly reduces the constraining power on the global bias parameters, which degrades the marginalised posteriors on $H_0$.
    
    \item The binning choices play a key role in the trade-off between signal sensitivity and shot noise in any tomographic analysis. The existing literature on cross-correlation of gravitational waves and galaxies use sensible binning choices adapted to the specific surveys under consideration. A useful rule of thumb is to choose bin widths that are significantly larger than the typical distance uncertainty of the gravitational wave events, while remaining sufficiently narrow to avoid excessive overlap between multiple redshift bins which would dilute the cross-correlation signal. To better understand how binning choices affect the cosmological inference, we examine their impact on the $H_0$ posteriors in Section~\ref{subsec:binning}. Figure~\ref{fig:binning} shows that, in the regime considered here, finer binning generally improves the constraints on $H_0$, with the most significant gains arising from increased resolution in redshift. However, this behaviour is strongly dependent on the number of gravitational wave events: as the event sample becomes sparse, shot noise in the gravitational wave field rapidly increases and can offset, or even outweigh, the benefits of finer binning.
\end{itemize}

Over the past decade, we have witnessed a surge of interest in the possibility of using gravitational wave observations in combination with galaxy catalogues for cosmological inference. \textcolor{black}{In the standard galaxy catalogue approach, the gravitational wave events are treated as independent and identically distributed (IID) draws from the underlying host galaxy population. By construction, this does not exploit the two-point clustering information between gravitational wave events. The cross-correlation between the gravitational wave and galaxy overdensity fields, as recently pointed out in \citet{Cheng:2026atn}, therefore represents a more general framework. Indeed, it contains the standard catalogue approach as the limiting case in which the prior assumptions about missing galaxies ignore the clustering information induced by the underlying cosmological model.} Furthermore, \citet{Cheng:2026atn} advocate including a population prior on the gravitational wave events, thereby combining the complementary information provided by the spectral-siren approach.

These theoretical advances, combined with promising forecasts for next-generation gravitational wave detectors, establish dark siren cosmology as a powerful and complementary approach for probing the Hubble tension. These prospects motivated us to investigate how a range of systematic biases and modelling choices, common to many large-scale structure cosmological analyses, might affect cosmological inference with this novel probe. Whereas previous studies in the literature have focused on the impact of different gravitational wave catalogues and uncertainty assumptions on the constraining power of the cross-correlation, in this work we deliberately assume optimal gravitational wave events and uncertainties in order to better isolate and understand the impact of pipeline choices and modelling strategies. Our results emphasise that the cross-correlation of dark sirens with galaxy redshift surveys is a powerful and complementary addition to the landscape of high-precision cosmological probes, but that it requires a consistent and carefully constructed modelling framework.

\section{Acknowledgments}
This research was conducted by the Australian Research Council through Discovery Project DP220101395 and the Centre of Excellence for Gravitational Wave Discovery (project number CE230100016).

\appendix
\section{A theoretical $S_{ij}$}
\label{app:theorysij}
In this section, we outline how the analytical $S_{ij}$ statistic can be computed. Starting from the expression for the angular cross-correlation in an $(i,j)$ bin (Equation~\ref{eq:Clij}), and substituting in the two projection kernels (Equations~\ref{eq:kernel1simp} and \ref{eq:kernel2simp}), we see that the resulting expression contains two integrals over redshift ($z$ and $z'$) and a single integral over wavenumber, $k$. The redshift integrals are coupled through products of spherical Bessel functions, $j_\ell(k\chi(z))j_\ell(k\chi(z'))$, which encode the radial mode projection.

Now consider the definition of $S_{ij}$ from Equation~\ref{eq:sij}. The key advantage of this definition is that, after summing over multipoles, the dependence on the spherical Bessel functions can be removed. Specifically, by invoking the angular closure relation for the spherical Bessel functions
\begin{equation}
\sum_{\ell=0}^{\ell_{\mathrm{max}}}(2\ell+1)j_{\ell}(k\chi(z))j_{\ell}(k\chi(z'))\approx \frac{\sin{k\left(\chi - \chi'\right)}}{k(\chi-\chi')}\;,
\end{equation}

where the approximation holds true as long as $\ell_{\mathrm{max}}$ is sufficiently large (the summation is only guaranteed to converge for $\ell_{\mathrm{max}}=\infty$).\footnote{``Sufficiently'' here depends on the nature of the kernel, but roughly speaking this approximation is valid so long as $\ell_{\mathrm{max}} \gg \chi / \Delta\chi$, where $\Delta\chi$ denotes the typical `resolution' we need in our measurement given the gravitational wave and galaxy distance/redshift errors and distance/redshift bins we are using i.e. $\Delta\chi \sim \sqrt{(\frac{\sigma_{d_{L}}}{(1+z)})^{2}+(\frac{c\sigma_{z}}{H(z)})^{2}+\frac{\Delta\chi_{i}^{2}}{12}+\frac{\Delta\chi_{j}^{2}}{12}}$.}

We can now define  

\begin{align}
    S_{ij}&\equiv \sum_{\ell=0}^{\ell_{\mathrm{max}}}\frac{\left(2\ell +1\right)}{4\pi} C_\ell^{\mathrm{ gal \times GW}}\left(i,j\right)\\
&\approx\frac{1}{N_iN_j}\int_{z_{i}^{\mathrm{min}}}^{z_{i}^{\mathrm{max}}}\int_{z_{j}^{\mathrm{min}}}^{z_{j}^{\mathrm{max}}} dz_1\; dz_2\; b_{\mathrm{gal}}(z_1)\frac{dn_{\mathrm{gal}}}{d z_1} b_{\mathrm{GW}}(z_2)\frac{dn_{\mathrm{GW}}}{d z_2}  \;\nonumber \\
&\qquad \times \int_0^\infty dk\;k^2 P(k) j_0\left(k\left[\chi_2-\chi_1\right]\right)\;,
\end{align}
where $N_{i,j}$ is the number of tracers in the bin considered, and $\chi_i \equiv \chi(z_i)$.
Let us briefly discuss the physical interpretation of the expression above. One can easily recognise in the integral over $k$ the definition of the 3-dimensional matter correlation function in real space, $\xi(r)$, with $r=|\chi(z_1) - \chi(z_2)$|. It follows that $S_{ij}$ corresponds to the correlation function evaluated at the comoving separation between the two bins, weighted by the respective window functions, i.e. the number densities of the tracers.

The physical interpretation of $S_{ij}$ is therefore straightforward. For a uniform sky distribution of sources, it reduces to a weighted measurement of the density correlation function along the line of sight. This differs from the estimator used in \citet{Ferri_2025}, namely the real-space angular correlation function (their Equation~2.6), with the two only coinciding in the limit of zero angular separation, $\theta = 0$. As shown in the main text, the loss of information from using this line-of-sight autocorrelation instead of the full angular correlation function is minimal. We therefore conclude that clustering along the line of sight alone is sufficient to obtain unbiased constraints on $H_0$.
\bibliographystyle{abbrvnat} 

\bibliography{bib.bib}




\end{document}